\documentstyle[twoside,fleqn,espcrc2]{article}

\newcommand{\AmS}{{\protect\the\textfont2
  A\kern-.1667em\lower.5ex\hbox{M}\kern-.125emS}}
\newcommand{\real}{{\sf I}\kern-.12em{\sf R}}

\title{The 3-loop Beta Function of QCD}

\author{B. All\'es\address{Dipartimento di Fisica and I.N.F.N.,
        Sezione di Pisa, Piazza Torricelli 2, 56126 - Pisa, Italy}
        ,
        C. Christou\address{Department of Natural Sciences, University
        of Cyprus, P.O. Box 537, Nicosia CY-1678, Cyprus}
        ,
        A. Feo$^{\rm a,}$\address{Scuola Normale Superiore, Pisa,
        Italy}
        ,
        H. Panagopoulos$^{\rm b}$\thanks{Presented the talk.} and
        E. Vicari$^{\rm a}$}

\begin{document}

\begin{abstract}
Using the background field technique, we calculate the 
3-loop beta function of lattice $SU(N)$ gauge theories.
In the pure gluonic case, we present our results, comparing to those
recently obtained by Luescher and Weisz. We also provide a progress
report in the case of QCD with Wilson fermions.
\end{abstract}

\maketitle

\section{INTRODUCTION}

The relationship between the bare coupling and the
cutoff is an essential ingredient in lattice calculations. 
Usual Monte Carlo simulations require the knowledge of this
relationship far from the critical point, where corrections to
asymptotic scaling could become relevant. 

In order to check the relevance of these scaling corrections,
we present here a calculation of the first non-universal
coefficient of the lattice $\beta$ function in pure Yang-Mills
theory with Wilson action \cite{afp}, 
as well as in the presence of Wilson fermions \cite{cfpv}.
This renormalization group function can be written as
\begin{equation}
\beta^L(g_0) \equiv - a \frac{\hbox{d}g_0}{\hbox{d}a} \mid_{g_r,\mu}=
- b_0 g_0^3 - b_1 g_0^5 - b_2^L g_0^7 - \cdots
\end{equation}
where $a$ is the lattice spacing, $g_0$ the lattice bare coupling,
$g_r$ is the renormalized coupling constant and $\mu$ the subtraction
point. It establishes how the bare coupling and the cutoff $a$ must 
simultaneously vary to keep renormalized quantities fixed.
The first two coefficients in eq. (1) are known as they are
scheme-independent. Our purpose is to compute the third coefficient,
$b_2^L$.

In a recent paper \cite{lw234} the authors calculated a quantity
related to this coefficient (in the pure gluon case),
using a coordinate space method \cite{lw429} 
for evaluating lattice integrals. In our calculation we have used
a different technique where superficially divergent integrands are
Taylor expanded in $D$ dimensions. The idea is a generalization 
to two loops of the procedure introduced in \cite{kawai}.
Thus, our results for the pure gluon case provide an independent check
of the calculation in reference \cite{lw234}.

In what follows we will first define the quantities we set out to
compute, and give a brief description of our calculational techniques.
We then present our results. We
discuss the effect of $b_2^L$ on the scaling behaviour of dimensionful
quantities. We close with some concluding remarks.

The involved algebra of lattice perturbation theory was
carried out by means of a computer code.
The main features of this code were outlined
in \cite{npb} and used
to compute the three loop perturbative background of the
topological susceptibility \cite{npb}, and the three loop lattice free
energy \cite{plb}, in pure Yang-Mills on the lattice.
 For the purpose of the present work, this code was extended to
include form factors, and to handle fermionic vertices.

Aside from the intrinsic interest of the result, this calculation
serves also as a prototype for numerous 2-loop calculations involving
fermions and external momenta, e.g. multiplicative renormalization of
fermionic currents. We expect to return to this issue in a future report.

\section{GENERAL CONSIDERATIONS}

We performed our calculation using the background field method. To
facilitate comparison, we have adopted the notation of
Ref.\cite{lw234}.

We want to calculate $\nu^{(2)}(p)$, defined through the lattice
background field two point function via:
\begin{equation}
\sum_\mu \Gamma^{(2,0,0)}(p,-p)_{\mu\mu}^{ab} = - \delta^{ab} 3 \hat
p^2 [1-\nu(p)]/g_0^2
\end{equation}
with $\nu(p) = \sum g_0^{2l}\nu^{(l)}(p)$. Knowledge of
$\nu^{(2)}(p)$, together with its $\overline{MS}$ counterpart
$\nu_R^{(2)}(p)$ and the one loop gluon self energy leads us, through
use of standard formulae, to $Z(g_0,\mu a) \quad (g_0 = Z(g_0,\mu a)
g(\mu))$. Knowledge of the three loop coefficient of the renormalized
$\beta$-function, $b_2^{\overline{MS}}$, along with the two loop coefficient $Z_{20}$ in the expansion:
\begin{equation}
Z(g_0,\mu a) = 1 + \sum_{i{=}1}\sum_{j{=}1}^i g_0^{2i} Z_{ij} \ln^j\mu
a
\end{equation}
gives us the three loop lattice $\beta$ function, through the relation:
\begin{equation}
b^L_2 = b_2^{\overline{MS}} - 2 b_1 Z_{10} + b_0 Z_{10}^2 + 2 b_0 Z_{20}
\end{equation}
All one loop quantities appearing above are well known. For the
continuum function $\nu_R^{(2)}(p)$, in the presence of $N_f$
fermionic species, we find (in the Feynman gauge):

\begin{equation} \nu_R^{(2)} (p) (16\pi^2)^2 = N^2 [8\rho
+577/18-6\zeta(3)]
\end{equation}
$$\,\, +N_f[(-3\rho-401/36)N+(\rho+55/12-4\zeta(3))/N] $$

\noindent
($\rho = \ln (\mu^2/p^2)$).

Contributions to $\nu^{(2)}(p)$ come from 35 diagrams in the pure
gluonic case (shown in Ref.\cite{lw234}). Fermions bring in 18 more
two-loop diagrams (the ghost diagrams of \cite{lw234}, with fermions
replacing ghosts) plus 2 one-loop diagrams and 2 diagrams with fermion
mass counterterms (we work with zero {\it renormalized} fermion mass).
We manipulate these diagrams using a computer code \cite{npb} which
has now been extended to include fermionic vertices and external
momenta. The code starts with a series of operations which can be
performed on diagrams with arbitrary number/type of
vertices/loops/external legs: Generation of vertices, contraction
with appropriate reassignment of indices, symmetrization, reduction of
Dirac and color indices, simplification of Lorentz structures. 
It is essential that the corresponding
algorithms be (quasi-)polynomial: The contrary results in astronomical
requirements of CPU and RAM, even in the simplest cases. This last
constraint turns out to be the most exacting one in the construction
of our algorithms.

At this point, we apply trigonometry judiciously to bring our
expressions into a ``canonical'' form, separating them also into (a
few) superficially divergent terms plus the remainder. This remainder
can be quite sizeable, especially in the fermionic case (often
thousands of terms) but it can be handled with a straightforward
Taylor expansion in external momenta or, when subdivergences are
present, with appropriate subtractions; e.g., in
\begin{equation}
(2\pi)^{-8}\int d^4q\,d^4k {\buildrel \circ \over q}\cdot {\buildrel
\circ\over k} \, {\buildrel \circ\over q}\cdot{\buildrel \circ\over p}
\, / (\hat q^2 \hat k^2 \widehat{q{+}k}^2 \widehat{k{+}p}^2)
\end{equation}
the subtraction: $1/\widehat{q{+}k}^2 = 1/\hat q^2 +
(1/\widehat{q{+}k}^2 - 1/\hat q^2)$ leads to a product of one loop
integrals with a single $\ln p$, plus a Taylor expandable
part. Different types of subtractions may be necessary for the full
evaluation of a diagram; suffice it to say that the most cumbersome
diagram (a ``diamond'' with a triangle of fermions) splits into ${\sim} 100$
{\it types} of expressions.

For the superficially divergent diagrams, we have extended beyond one loop
the proceduce of Ref. \cite{kawai} (analytic continuation to
$4{-}2\epsilon$ dimensions, followed by Taylor expansion). 
This entails a lengthy series of additions/subtractions
to parts of the integrand and to the integration domain (so that, e.g., 
$1/\widehat{k{+}q}^2 = 1/\hat q^2 + \ldots$, $1/\hat q^2 = 1/q^2 + \ldots$, 
$[{-}\pi/a,\pi/a]^4 = R^4 - \ldots$, etc.). Interchanging the
$\epsilon{\to}0$ and $a{\to}0$ limits is more complicated in this
case. Also, certain intermediate
expressions must be calculated to ${\cal O}(\epsilon)$. 
Proceeding in this way we have produced \cite{afp} a table of values
for all such two-loop diagrams, complementing those given in
\cite{lw234}.

Our numerical integrations were done on finite lattices, with
subsequent extrapolation to infinity, using the form $\sum_{m,n}
a_{mn} (\ln^mL)/L^n$ ($L$: lattice size). This leads typically to four
significant digits. The coordinate space method of Ref.\cite{lw429}
gives a much better precision, but cannot be applied as is to
fermionic diagrams.

\vfill
\section{RESULTS}

The contribution $\nu_i(p)$ of the $i$th diagram to $\nu^{(2)}(p)$ is:
$$\hat p^2 \nu_i(p) = c_{0,i} + c_{1,i} \sum_{\mu{=}0}^3 {p_\mu^4\over
p^2} + p^2 \big\{ c_{2,i} [\ln p^2/(4 \pi)^2]^2$$
\begin{equation}
\quad +c_{3,i} \ln p^2
/(4\pi)^2 + c_{4,i}\big\} + {\cal O}(p^4) 
\end{equation}
where: $c_{n,i} = c_{n,i}^{(0)}/N^2 + c_{n,i}^{(1)} + c_{n,i}^{(2)}
N^2$ for diagrams without fermions and  $c_{n,i} = c_{n,i}^{(0)}/N +
c_{n,i}^{(1)}N$ for diagrams with fermions. 

Gauge invariance requires that $\sum_i c_{0,i} = 0$; this is indeed
so, separately for diagrams with and without fermions, as we have
checked both by numerical evaluation and algebraically (expressing all
vertices involved in terms of derivatives of the gluon and fermion
propagators, and performing integration by parts where
necessary). Lorentz invariance requires: $\sum_i c_{1,i} = 0$. Such
contributions appear in non-tadpole diagrams containing one-loop
renormalized gluon propagators; again, we verified that their sum
vanishes, both numerically and algebraically. The coefficients
$c_{2,i}$  of the logarithms squared must coincide with those of the
corresponding continuum diagrams and indeed they do. 

In the pure gluonic case, we compared our results to those of
L\"uscher and Weisz. All coefficients $c_{0,i}, c_{1,i}, c_{2,i},
c_{3,i}$ coincide. As for $c_{4,i}$, the parts $c_{4,i}^{(0)}$ and
$c_{4,i}^{(1)}$ agree for each diagram. The coefficients
$c_{4,i}^{(2)}$ agree for all but 1 diagram (\# 5 of
\cite{lw234}), leading to a slightly smaller (${<}3\%$) total for
$c_4^{(2)}$. To this writing, we have not been able to find the
cause of this discrepancy. 

Thus, comparing with the expression of \cite{lw234} for $\nu^{(2)}(p)$
without fermions:
\begin{eqnarray}
\nu^{(2)}(p) &=& - N^2 /(32\pi^4) \ln p^2 + 1/N^2 (3/128)\nonumber \\
&& -0.01654462 + N^2 0.0074438722
\label{nu2}\end{eqnarray}
we find for the last number in the above: 0.007230.

Our results for $\nu^{(2)}(p)$ with Wilson fermions are in the final
stage of extrapolation/precision checks \cite{cfpv}. 

\section{CONCLUSIONS}

Eq.(\ref{nu2}) for $\nu^{(2)}(p)$ can be used directly to
evaluate $b_2^L$, as explained. In the absence of fermions, we find: 
\begin{equation}
(16 \pi^2)^3 b_2^L = -2143/N + 1433.8 N - 366.2 N^3
\end{equation}
Substituting now in the asymptotic scaling formula: 
\begin{eqnarray}
a \Lambda_L &=& \exp(-1/2b_0g_0^2) (b_0g_0^2)^{-b_1/2b_0^2}\cdot \\
&& (1 + g_0^2
(b_1^2 - b_2^L b_0)/2b_0^3 + \ldots) \nonumber
\end{eqnarray}
the last factor becomes, for $SU(3)$:
\begin{equation}
(1 + 0.1896 g_0^2) 
\end{equation}
Thus, in the region $g_0\sim 1$ this factor
brings about a rather sizeable correction to asymptotic scaling. 

Some further computations which can be carried out without changes in
our computer code are two loop renormalization of various composite
operators (fermionic currents, etc.), use of improved actions and
(with some modifications) use of staggered fermions.

\end{document}